\documentclass[aps,preprint,preprintnumbers, amsmath, amssymb, prb,showpacs]{revtex4-1}
\usepackage{times}
\usepackage{graphicx}
\usepackage{psfrag}
\usepackage{ae}
\usepackage{amsmath,amssymb}
\usepackage[usenames]{color}
\usepackage{float}

\begin{document}
\author{Leandro B. Krott} 
\email{leandro.krott@ufrgs.br}
\affiliation{Programa de P\'os-Gradua\c c\~ao em F\'isica, Instituto 
de F\'{\i}sica, Universidade Federal
do Rio Grande do Sul, Caixa Postal 15051, CEP 91501-970, 
Porto Alegre, RS, Brazil}

\author{Jos\'e Rafael Bordin} 
\email{josebordin@unipampa.edu.br}
\affiliation{Campus Ca\c capava do Sul, Universidade Federal
do Pampa, Av. Pedro Anuncia\c c\~ao, s/n, CEP 96570-000, 
Ca\c capava do Sul, RS, Brazil}

\author{Marcia C. Barbosa} 
\email{marciabarbosa@ufrgs.br}
\affiliation{Instituto 
de F\'{\i}sica, Universidade Federal
do Rio Grande do Sul, Caixa Postal 15051, CEP 91501-970, 
Porto Alegre, RS, Brazil}

\title{New structural anomaly induced by nanoconfinement}

 \date{\today}

\begin{abstract}

We explore the structural properties of anomalous fluids confined in a nanopore
using Molecular Dynamics simulations. The fluid is modeled by
core-softened (CS) potentials that have a repulsive shoulder and an 
attractive well at a further distance. Changing the attractive well depth of the 
fluid-fluid interaction potential, we studied the behavior of the anomalies
in the translational order parameter $t$ and excess entropy $s_{ex}$
for the particles near to the nanopore wall (contact layer) for systems with two 
or three layers of particles. When the attractive well of the CS potential 
is shallow, the systems present a three to two layers transition and, additionally
to the usual structural anomaly, a new anomalous region in $t$ and $s_{ex}$. 
For attractive well deep enough, the systems change from three layers 
to a bulk-like profile and just one region of anomaly in $t$ and $s_{ex}$
is observed. Our results are discussed in the basis of the fluid-fluid 
and fluid-surface interactions.

\end{abstract}


\maketitle
\section{Introduction}

Anomalous fluids exhibit a set of properties called anomalies that divert from the observed
in simple fluids. The increase of density with the temperature at a fixed pressure
and the increase of diffusivity under compression are examples of 
these anomalies. Water is the most well known fluid that
present thermodynamic, dynamic and structural anomalous 
behavior~\cite{Ke75,An76,Pr87}, with 70 known anomalies~\cite{URL}. 
In addition, Te~\cite{Th76}, Bi~\cite{Handbook}, 
Si~\cite{Sa67,Ke83}, $Ge_{15}Te_{85}$~\cite{Ts91},  liquid 
metals~\cite{Cu81}, graphite~\cite{To97}, silica~\cite{An00,Sh02,Sh06}, silicon~\cite{Sa03},  
$BeF_2$~\cite{An00} exhibit thermodynamic anomalies~\cite{Pr87} while
silica~\cite{Sa03,Sh02,Sh06,Ch06} and silicon~\cite{Mo05} 
show a maximum in the diffusion coefficient at constant temperature, similar to
what happens in water~\cite{Ne01, Ne02a}.

What these materials have in common that allow them to have
anomalous properties? In order to answer to this question a 
measure of order of the liquid system was proposed~~\cite{Tr00,Er01,Er03,Sh02}. The translational order parameter, 
\begin{equation}
t = \int_{}^{}\mid g(r)-1\mid d^3r
\end{equation}
measures the tendency of pairs of molecules to adopt a preferential separation. $t$  
vanishes for an ideal gas, and is large for a crystal. For normal liquids $t$ increases with the increase with density, since
the liquid becomes more ordered as it becomes more dense. In anomalous
liquids there is a region in density in which $t$ decreases with 
the increase of density. This decrease of structure with
the increase in density indicates that the pairs
of particles have two preferential separations: one 
more ordered 
in which particles are at further distance and 
another more disordered in which particles are closer~\cite{Er01}.
As the density increases these bimodal distribution changes 
favoring the more disordered structure.

While $t$ indicates through the structure 
the presence of the length scales the excess entropy~\cite{Ne58,Ra71,Wa87,Ba89}
gives a thermodynamic measure  of the presence of 
anomalies in liquids~\cite{Er06}. 
The excess entropy is
defined as the difference between the entropy of a 
real fluid and that of an ideal gas at
the same temperature and density, namely $s_{ex}=s-s_{ig}$.
In principle the ideal entropy is $s_{ig}=\ln{\rho}+f(T)$ where
$f(T)$ is a function  of the temperature only. In the 
limit of  $\rho\rightarrow \infty$, $s_{ex}\rightarrow 0_{-}$.
For normal liquids as  the density increases, $s_{ex}$ decreases.
 The anomalous materials described  above are  characterized for having a 
region in density in which $(\partial s_{ex}/\partial \ln{\rho})>0$~\cite{Er01,Er06}. This unusual behavior of 
$s_{ex}$ is also related to the presence
of two length scales~\cite{Er06}. Even though $t$ is related with the structure while $s_{ex}$ to the 
thermodynamics these two quantities are in fact related. This becomes
clear if the two body approximation for $s_{ex}$ namely 
\begin{equation}
s_2 = \rho\int_{}^{} [g(r)\ln g(r) - g(r) +1]d^3r
\end{equation}
is employed. From the definitions of $s_2$ and $t$ , both of which depend
on deviations of $g(r)$ from unity, it is to be expected that
variations in $s_2$ and $t$ would be anticorrelated. Thus, the existence 
of a maximum in $s_2$ at high 
densities implies a minimum in $t$ as a function of density.
These two quantities connect the thermodynamic
and structure by the two length scales.

Notwithstanding the relevance of the thermodynamic and 
dynamic anomalous bulk properties of the systems above, novel
developments had arisen in the confined structures~\cite{Tabeling14}.
In confined systems, crystallization is not uniform and depends on the 
size of the nanopores~\cite{De10,Ja08}. Simulations for SPC/E water, 
for example, show partial crystallization inside nanotubes that 
leads to phase transitions not observed in bulk system~\cite{Koga01,
MaG97}. A transition between a bilayer ice and a trilayer fluid for 
different degrees of confinement also are observed for water
~\cite{Giovambattista09,Lo09}. Under high degrees of confinement, 
water can form a monolayer ice and behaves very different from bulk 
systems~\cite{Zangi03a,Santos12}, similar to what happens with two-dimensional 
core-softened fluids~\cite{Du14a,Du14b,Du14c}. For confined systems,
some anomalous liquids also present layering transition~\cite{Bordin14a}, 
superflow~\cite{Jakobtorweihen05,Qin11, Bordin13a, Bordin14b} and distinct dynamic 
behavior~\cite{Krott13b,Krott13a,Bordin12b}. Oscillations in the 
solvation  force~\cite{Ho81} and a dramatic increase of the 
viscosity can occur in ultrathin confined fluids~\cite{Gr91}.

What does drive these novel phenomena observed in nanoconfined anomalous fluids?
Under confinement, anomalous 
fluids exhibit properties not observed in bulk~\cite{Koga97,
Koga01,St12,Ja08,Fa09,De10,Er11}.
The confined fluid
is not distributed uniformly in the nanopore but
forms layers. Therefore, the new
anomalous properties that arise under confinement
are related to the presence and structure of
the layers. For instance, each layer might crystallize at 
a different temperature~\cite{Krott14a}. Also the number of layers and their
structures depend on the nanopore size and structure and
on the fluid-wall interaction potential~\cite{Bordin12b,Krott13a}.  
Acknowledging that the presence of the layering structure is 
responsible for the novel behavior
observed in anomalous fluids under confinement it is reasonable to 
think that the new properties  appear as the result
of the competition between the two fluid-fluid  length scales 
and the fluid-wall length scale.

In order to check this hypothesis, in this paper we explore the 
behavior of the translational order parameter, $t$, and 
the excess entropy, $s_{ex}$, as a 
function of density and temperature of a confined model system of
particle interacting through a core-softened potential.
This two length scales coarse-grained potential in the bulk 
exhibits the density, the diffusion and the structural
anomalies~\cite{Oliveira06a,Oliveira06b,Silva10} observed in the water-like systems 
listed above. Under confinement this potential shows   
the formation of layers~\cite{Krott13a,Krott13b,Krott14a,Bordin12b,Bordin13a,Bordin14b}.
Here we test for different ratios between the two fluid-fluid length scales
if the presence of new anomalies in $t$ and in $s_{ex}$ are 
associated with changes in the layering structure. Our results
give support to the surmise that the anomalies appear as 
the result of competition between bonding, nonbonding, hydrophilic 
and hydrophobic interactions.

The paper is organized as follows: in Sec. II we introduce the model and 
describe the methods and simulation details; the results are given
and discussed in Sec. III; and in Sec. IV we present our conclusions.

\section{The Model and the Simulation details}
\label{Model}

\subsection{The Model}

\begin{figure}[ht]
\begin{center}
\includegraphics[clip=true,width=8cm]{fig1a.eps}
\includegraphics[clip=true,width=8cm]{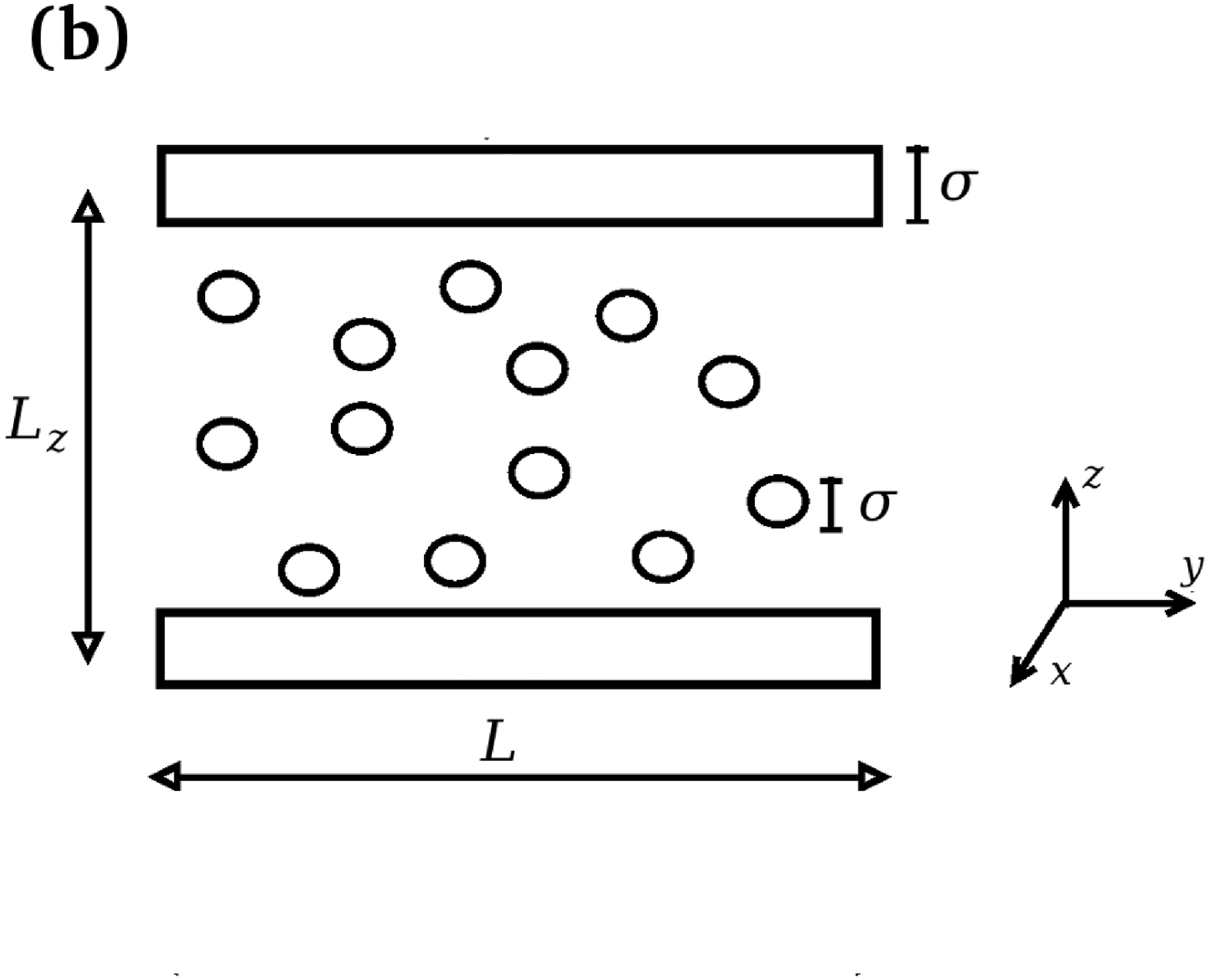}
\end{center}
\caption{(a) Particle-particle interaction potentials given by Eq.~(\ref{alanE})
with parameters $a = 5.0$, $r_0/\sigma = 0.7$, $c = 1.0$, $r_1/\sigma = 3.0$ 
and $d = 0.5$, for different values of $b$. (b) Schematic depiction of the fluid confined 
between two smooth walls.}
\label{fig2}
\end{figure}

The anomalous fluid was modeled using an isotropic effective potential~\cite{Silva10} given by 

\begin{equation}
\label{alanE}
\frac{U(r_{ij})}{\varepsilon} = 4\left[ \left(\frac{\sigma}{r_{ij}}\right)^{12} -
\left(\frac{\sigma}{r_{ij}}\right)^6 \right] + a\exp\left[-\frac{1}{c^2}\left(\frac{r_{ij}-r_0}{\sigma}\right)^2\right]
-b\exp\left[-\frac{1}{d^2}\left(\frac{r_{ij}-r_1}{\sigma}\right)^2\right],\nonumber
\end{equation}
 
\noindent where $r_{ij} = |\vec r_i - \vec r_j|$ is the distance between the two fluid particles $i$ and $j$.
The first term of this equation is a standard 12-6 Lennard-Jones (LJ) potential~\cite{AllenTild}.
The second and third therms are Gaussians centered at $r_0$ and $r_1$, with depth $a$ and $b$ and width $c$
and $d$, respectively. The fixed parameters of Eq.~(\ref{alanE}) are: 
$a = 5.0$, $r_0/\sigma = 0.7$, $c = 1.0$, $r_1/\sigma = 3.0$ 
and $d = 0.5$. Changing the parameter $b$, the attractive part increases without 
change the repulsive shoulder at $r\approx 1.2$.
For $b = 0$ the potential is purely repulsive and 
presents density, diffusion and structural anomalies in bulk~\cite{Oliveira06a, Oliveira06b} 
and in confined systems~\cite{Bordin12b, Bordin13a, Krott13a, Krott13b, Krott14a, Bordin14a, Bordin14b}.
Increasing $b$, beyond these anomalies, gas-liquid and liquid-liquid critical points appear in bulk 
systems~\cite{Silva10}. Besides $b = 0$ (model A), the potentials studied correspond
to $b = 0.25$ (B), $0.50$ (C) and $0.75$ (D), as illustrated in Fig.~\ref{fig2} (a). 

\begin{figure}[ht]
\begin{center}
\includegraphics[clip=true,width=10cm]{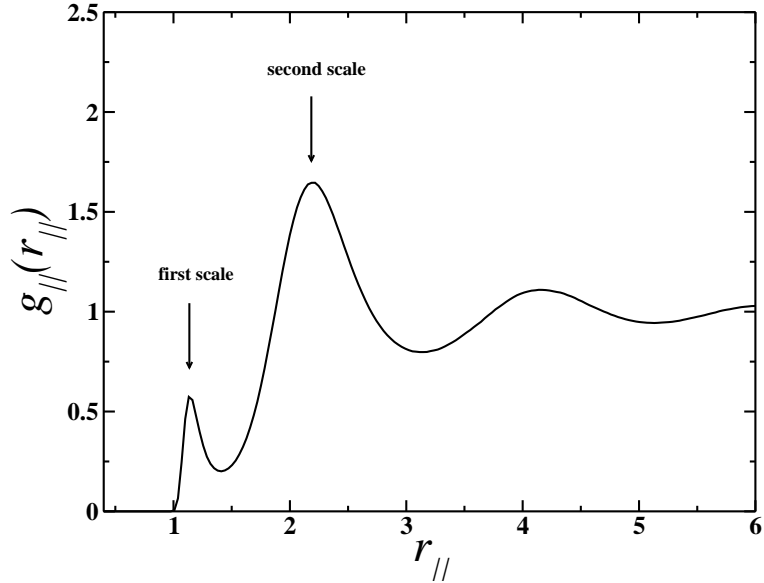}
\end{center}
\caption{Radial distribution function for potential Model A for the bulk 
system at $\rho = 0.137$ and $T = 0.175$. The two peaks represent the two fluid-fluid interaction length
scales, as indicated by arrows.}
\label{gr}
\end{figure}

All these potentials are characterized by two length scales: one at
the shoulder distance and another at the minimum of the potential.
These two length scales can be seen more explicitly in the 
bulk radial distribution function that exhibits two peaks at these
two representative distances~\cite{Oliveira06b} as illustrated in Fig.~\ref{gr}.

In all cases the system was composed by $N$ spherical particles of diameter $\sigma$ and 
mass $m$ confined between two smooth fixed walls, or plates, with area $L^2$.
The center-to-center plates distance is $L_z$. 
A schematic depiction of the system is shown in Fig.~\ref{fig2} (b).
The particle-wall interaction was given by the sixth power (R6) 
potential~\cite{Krott14a}, 
 
\begin{equation}\label{eq_potentialr6}
U_{R6} = \left\{ \begin{array}{ll}
 4\varepsilon\left(\sigma/z_{ij} \right)^{6} + 0.1875\varepsilon\left(z_{ij}/\sigma\right) -
U_{R6c}  , \qquad z_{ij} \le z_{cw} \\
0   , \qquad z_{ij}  > z_{cw} \;,
\end{array} \right.
\end{equation}

\noindent where $z_{cw} = 2.0\sigma$ and $U_{R6c}  
= 4\varepsilon\left( \sigma/z_{cw} \right)^{6}+ 0.1875\varepsilon(z_{cw}/\sigma)$. 
The term $z_{ij}$ measures the distance between the wall 
at $j$ position and the $z$-coordinate of the fluid particle $i$. This potential 
represents a hydrophobic nanopore-fluid interaction.

\subsection{The simulation details}

The simulations were performed in the $NVT$ ensemble considering $N = 507$ particles. The plates 
have fixed positions and the distances between them was varied 
from $L_z = 5.3\sigma$ to $L_z = 7.5\sigma$, depending on the model considered. For each 
system, at a fixed $L_z$, different densities were obtained changing 
the simulation box size in the $x$ and $y$ direction, $L$,
and consequently the plates size, from $20\sigma$ to $65\sigma$. 
Standard periodic boundary conditions were applied in the $x$ and $y$ 
directions. Because of the excluded volume due the fluid-plate
interaction, the distance $L_z$ between the plates needs to be corrected 
to an effective distance~\cite{Ku05, kumar07} that can be approached by $L_{ze} \approx L_z -\sigma$.
The effective density will be $\rho_e = N/(L_{ze}L^2)$. The symbol $_e$ will be 
omitted in order to simplify the discussion.

The velocity Verlet algorithm was used to integrate the
equations of motion for the fluid particles, considering a time step 
of $\delta t = 0.001$ in LJ units. We performed $4\times10^5$ steps to equilibrate the system and 
$8\times10^5$ steps to obtain the physical quantities. The temperature was 
kept fixed through the Nose-Hoover heat-bath with a coupling parameter $Q = 2$.
The temperatures studied were different for each model considered: 
$k_BT/\varepsilon = 0.150$, $0.250$ and $0.400$ for the model A; 
$k_BT/\varepsilon = 0.200$, $0.300$ and $0.500$ for the model B; 
$k_BT/\varepsilon= 0.300$ and $0.500$ for the model C; and 
$k_BT/\varepsilon = 0.500$ and $0.600$ for the model D.
The temperatures, densities and separation of plates were chosen
according to the particularities of each model~\cite{Silva10}. 
The fluid-fluid interaction (Eq.~(\ref{alanE})), has a 
cutoff radius $r_{\rm c}/\sigma = 4.5$ for all models.

We analyze the structure of the system using the lateral radial distribution 
function $g_{||}(r_{||})$ and the translational order parameter, $t$. The 
$g_{||}(r_{||})$ is defined as

\begin{equation}
\label{gr_lateral}
g_{||}(r_{||}) \equiv \frac{1}{\rho ^2V}
\sum_{i\neq j} \delta (r-r_{ij}) \left [ \theta\left( \left|z_i-z_j\right| 
\right) - \theta\left(\left|z_i-z_j\right|-\delta z\right) \right],
\end{equation}
where the Heaviside function $\theta (x)$ restricts the sum of particle pair in a 
slab of thickness $\delta z = \sigma$ for the contact layer. The radial
distribution function is proportional to the probability of finding a particle
at a distance $r$ from a referent particle.

The translational order parameter $t$ is defined as \cite{Sh02,Er03,Er01}
\begin{equation}
\label{order_parameter}
t \equiv \int^{\xi _c}_0  \mid g_{\parallel}(\xi)-1  \mid d\xi,
\end{equation}
\noindent where $\xi = r_{\parallel}(\rho^{l})^{1/2}$ is the interparticle 
distance in the direction parallel to the plates scaled by the density of the layer,
$\rho^{l} = N^{l}/L^2$. $N^{l}$ is the average of particles for each layer. 
We use $\xi_c = (\rho^{l})^{1/2}L/2$ as cutoff distance. The parameter $t$ measures
how structured is the system. For an ideal gas, $g(r) = 1$ and, consequently, $t = 0$,
whereas for structured phases, like crystal and amorphous solids, $t$ can assume 
large values.

The excess entropy is defined as the difference between the entropy of a real fluid and the ideal
gas at the same temperature and density. As the systems are organized in layers
of different structures, it is possible to define an excess entropy 
for each one, that can be approached as follows~\cite{Ne58,Ra71,Wa87,Ba89}

\begin{equation}
\label{excess_eq}
s_{ex} \approx -2\pi \rho^{l} \int^{\infty}_{0}\left[ g_{||}(r_{||}) 
\ln g_{||}(r_{||})-g_{||}(r_{||})+1\right]r_{||}^2dr_{||}. 
\end{equation}

The physical quantities will be measured in the standard LJ units~\cite{AllenTild},
namely
\begin{equation}
\label{red1}
r^*\equiv \frac{r}{\sigma}\; \quad 
\mbox{and}\quad \rho^{*}\equiv \rho \sigma^{3}\;,
\end{equation}
for distance and density of particles, respectively, and
\begin{equation}
\label{rad2}
T^{*}\equiv \frac{k_{B}T}{\epsilon}\quad \mbox{and}\quad s_{ex}^*\equiv \frac{s_{ex}}{k_B} 
\end{equation}
for temperature and excess entropy, respectively. 
Since all physical quantities are defined in reduced LJ units in this paper, 
the $^*$ will be omitted, in order to simplify the discussion.

Data errors are smaller than the data points and are not shown. The 
data obtained in the equilibration period was not considered for the quantities evaluation.

\section{Results and Discussion}
\label{Results}

Usually fluids confined between flat plates are structured in layers.
They can be classified  in contact layers, which are in contact
with the  walls, and central layers, which are in the nanopore center
without contact with the walls.
The layer properties depend on the temperature, density and separation of the plates.
In order to relate the fluid anomalies with the
structure of the layers we have analyzed systems with 
two or three layers.

\begin{figure}[!htb]
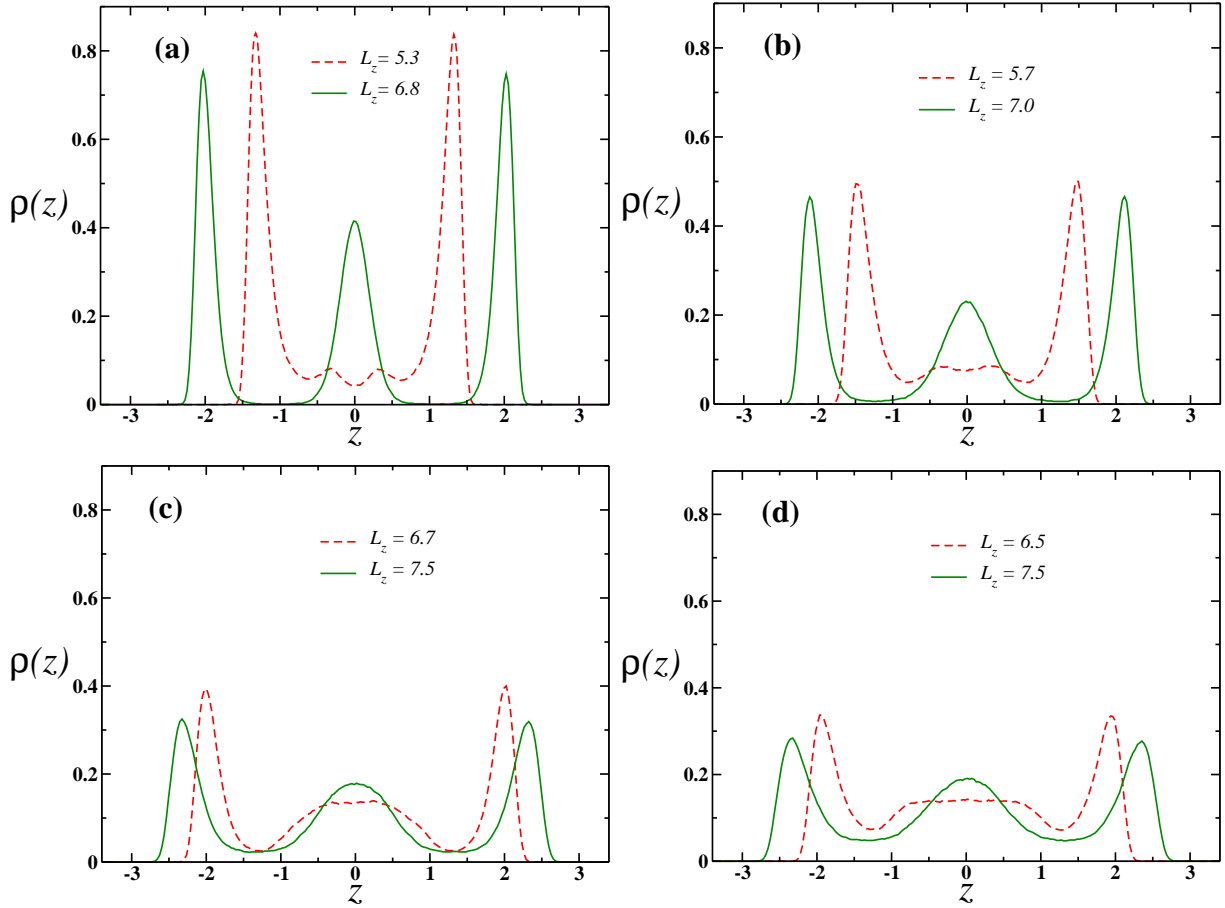

 \centering\begin{center}
 \begin{tabular}{cc}
\includegraphics[clip=true,width=8cm]{fig3a.eps}
\includegraphics[clip=true,width=8cm]{fig3b.eps}\\
\includegraphics[clip=true,width=8cm]{fig3c.eps}
\includegraphics[clip=true,width=8cm]{fig3d.eps}
\tabularnewline
 \end{tabular}\par
\end{center}
\caption{Transversal density profiles for large and narrow systems at lowest temperature
for (a) model A at $T = 0.150$, (b) model B at $T = 0.200$, (c) model C at 
$T = 0.300$ and (d) model D at $T = 0.500$. The correspondent densities for each 
case are given in the text.}
\label{histograms}
\end{figure}

The Fig.~\ref{histograms} shows the 
transversal density profiles for each model. The Fig.~\ref{histograms}(a) 
shows that at $T = 0.150$ potential A shows
a transition from a regime of three layers at  $L_z=6.8$ and $\rho = 0.111$
to a regime of two layers for at $L_z = 5.3$ and  $\rho = 0.150$.  
Fig.~\ref{histograms}(b) shows a similar behavior for the potential B 
at $T = 0.200$ that exhibits three layers for  $7.0$ and  
$\rho = 0.108$ and two layers for  $L_z = 5.7$ and  $\rho = 0.137$.
Fig.~\ref{histograms}(c) for the potential C at $T = 0.300$ shows a 
different behavior. For both $L_z = 6.7$ with  $\rho = 0.113$ and $7.5$
with $\rho = 0.095$ between the two contact layers there is continuous 
distribution of particles forming an interlayer. Fig.~\ref{histograms}(d) 
for the potential D at $T = 0.500$ shows also no transition
when the confining distance changes from $L_z=7.5$ and $\rho = 0.099$ 
to $L_z = 6.5$
and  $\rho = 0.117$.

Figs.~\ref{histograms} show that in addition to the
confining distance the well depth in Eq.~(\ref{alanE}) 
plays an important role in the number and structure  of fluid layers. For the 
pure repulsive case, model A, the system shows distinct layers since particles in two 
different layers have no attraction. In addition the transition from two to three layers 
happens when a new layer can be accommodated satisfying the minimum of the fluid-fluid and wall-fluid energies.
The model B, with a shallow well, shows a very similar behavior. On the 
other hand, for the cases with deep attractive well, models C and D, the 
system exhibits three not well defined 
layers for both plate separations. The central layer is present for distances 
between the confining walls that would imply that the fluid-fluid distance 
between particles in two neighbor layers is smaller than the minimum of 
the interparticle potential. The competition between the confinement and
the fluid attraction leads to this scenario.

In order to test this in more detail in the next section
the translational order parameter and the excess entropy
will be analyzed for the four potentials.

\subsection*{Translational order parameter}

\begin{figure}[!htb]
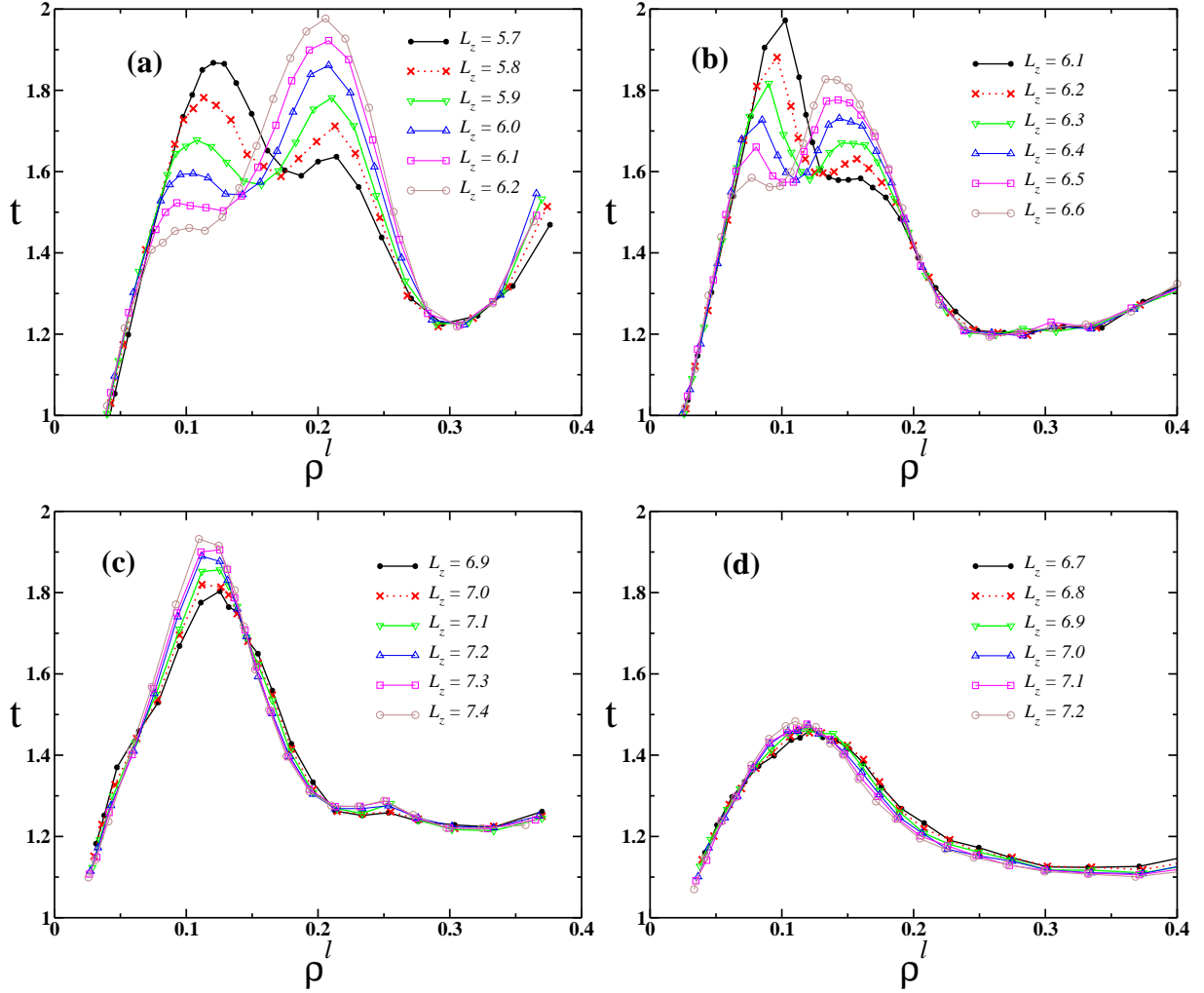

 \centering\begin{center}
 \begin{tabular}{cc}
\includegraphics[clip=true,width=8cm]{fig4a.eps}
\includegraphics[clip=true,width=8cm]{fig4b.eps}\\
\includegraphics[clip=true,width=8cm]{fig4c.eps}
\includegraphics[clip=true,width=8cm]{fig4d.eps}
\tabularnewline
 \end{tabular}\par
\end{center}
\caption{Translational order parameter $t$ as function of layer density for 
(a) model A at $T = 0.150$, (b) model B at $T = 0.200$, (c) model C at 
$T = 0.300$ and 
(d) model D  at $T = 0.500$. The other temperatures and separation of plates 
were not shown for simplicity.}
\label{pt}
\end{figure}

The translational order parameter, $t$,
was measured for the contact layer according to 
Eq.~\ref{order_parameter} for all the four models. Fig.~\ref{pt}
shows the parameter $t$ as function of layer density $\rho^l$ for (a) model A,
(b) model B, (c) model C and (d) model D for fixed distances between
the walls.

In normal fluids $t$ increases monotonically with $\rho$ for 
all temperatures, but 
anomalous fluids exhibit a region in the pressure 
versus temperature phase diagram in which $t$ decreases
with $\rho$ leading to a density of maximum $t$ at $\rho_{tmax}$ and 
a density of minimum $t$ at $\rho_{tmin}$. The interval
of densities between $\rho_{tmax}<\rho<\rho_{tmin}$ 
defines the anomalous region in the 
pressure versus temperature phase diagram. The 
two densities, $\rho_{tmax}$ and $\rho_{tmin}$, are associated with
the two characteristic
length scales of potential: one close 
to $r\approx 1.2$ and 
another at $r\approx 2.5$. 

In the case of confined systems  Fig.~\ref{pt} (c) and (d)
shows that the cases of the deepest
attractive part, models C and D, exhibit
the same anomalous behavior as the bulk
with a layer density of maximum $t$, $\rho^l_{tmax}$,
and a density of minimum $t$,  $\rho^l_{tmin}$.
The cases A and B, however, show an 
additional density of maximum  $t$,  $\rho^l_{tmax1}$, and 
density of minimum $t$,  $\rho^l_{tmin1}$,
that can not be associated with the two length scales
of the bulk system.

\begin{figure}[!htb]
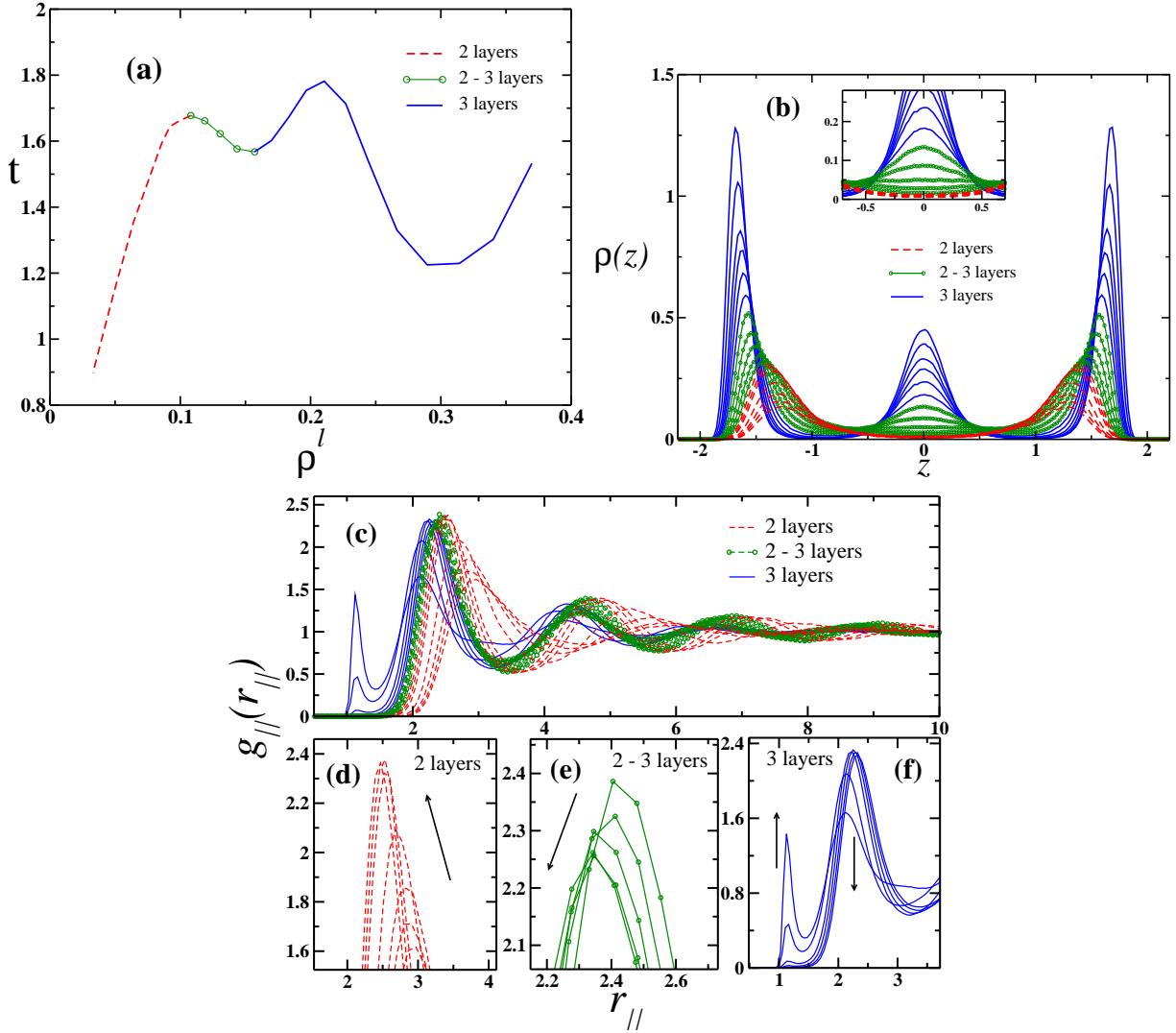

 \centering
 \begin{tabular}{cc}
\includegraphics[clip=true,width=8cm]{fig5a.eps}
\includegraphics[clip=true,width=8cm]{fig5b.eps}\\
\includegraphics[clip=true,width=10cm]{fig5c.eps}
\tabularnewline
 \end{tabular}\par
 \caption{Model A with plates separated by $L_z = 5.9$, temperature
$T = 0.150$ and densities $\rho^l < 0.108$ (red dashed line), $0.108 \le \rho^l \le 0.157$
(green dotted line) and $\rho^l > 0.157$ (blue solid line). 
In (a) we have the translational order parameter as function 
of layer density, in (b) the transversal density profiles, in (c) the lateral radial 
distribution function (LRDF) versus lateral distance, in (d) and (e) a zoom of the first 
peak of the LRDF for systems with two layers and two to three layers, respectively, and in 
(f) the competition between scales in the LRDF for systems with three layers. The arrows indicate 
the increase of density.}
\label{MA_t}
\end{figure}

Which mechanism leads to this new region of structural anomaly? 
In order to check if the new anomalous region
is related to new structural arrangements not present
in the bulk system, the densities of 
maximum and minimum $t$, 
$\rho^l_{tmax1}, \rho^l_{tmin1}, \rho^l_{tmax}$ and $ \rho^l_{tmin} $ were 
inspected for number of layers.

Fig.~\ref{MA_t} (a)
shows the translational order parameter as function of layer density
for plates separated for $L_z = 5.9$ and temperature $T = 0.150$. 
In addition to the expected minimum 
of $t$ in $\rho^l_{tmin} \approx 0.3$ (also observed in bulk system), 
a second minimum around $\rho^l_{tmin1} \approx 0.157$ appears.
Likewise, besides the expected maximum, of 
$t$ at  $\rho^l_{tmax} \approx 0.2$, another maximum at  
$\rho^l_{tmax1} \approx 0.108$ appears.

The colors in the Fig.~\ref{MA_t} (a) identify the number of layers in
each region of the $t$ versus contact layer density plot.
Fig.~\ref{MA_t} (b) shows that for high 
densities, $\rho^l > \rho^l_{tmin1}$ (blue curves),
the system is ordered in
three layers, while for low densities,  $\rho^l < \rho^l_{tmax1}$ 
(red curves), shows just two well defined layers. 
The inset shows a zoom in the center of the plates
illustrating that the new anomalous
region in $t$ happens for $\rho^l_{max1}<\rho^l<\rho^l_{min1}$, the
region of densities where the 
system melts the central layer. 

For the bulk system the peaks
at the radial distribution function can
be associated with the anomalous behavior of $t$ and $s_{ex}$. 
For a fixed temperature two peaks associated 
with the two length scales of the potential at the $g(r)$ are present.
As the density increases, the peak in the $g(r)$ associated with  the 
smaller length scale, $r\approx 1.2$,  increases  
while the larger peak, $r\approx 2.5$,  decreases. The same phenomena 
can be seen for the lateral radial distribution function versus the lateral 
distance for $\rho^l>\rho^l_{min1}$  illustrated in 
Fig.~\ref{MA_t} (c) (the blue plots) and Fig.~\ref{MA_t} (f). This
means that particles in all the three layers accommodate in
arrangements in the two length scales.  For
low densities, $\rho^l<\rho^l_{tmax1}$, 
there is only a peak at the larger length scale and this peak
increases with the increase of density as illustrated in Fig.~\ref{MA_t} (c)
(the red plots)
and Fig.~\ref{MA_t} (d) what is also observed in
normal bulk systems. This implies
that particles are arranged in the further length scale. For $\rho^l_{tmax1}<\rho^l<\rho^l_{tmin1}$  
the behavior, shown in Fig.~\ref{MA_t} (e), is different from the observed
for bulk systems. The decrease in density implies an
increase in the peak of the larger length scales because
as the density is decreased the central layer melts 
and more particles are present in the contact layer, increasing
the number of particles at the further length scale. The 
particular density in which the transition happens is related 
to the confining distance imposed by the wall-fluid interaction
and the minimum of the fluid-fluid potential.

\begin{figure}[!htb]
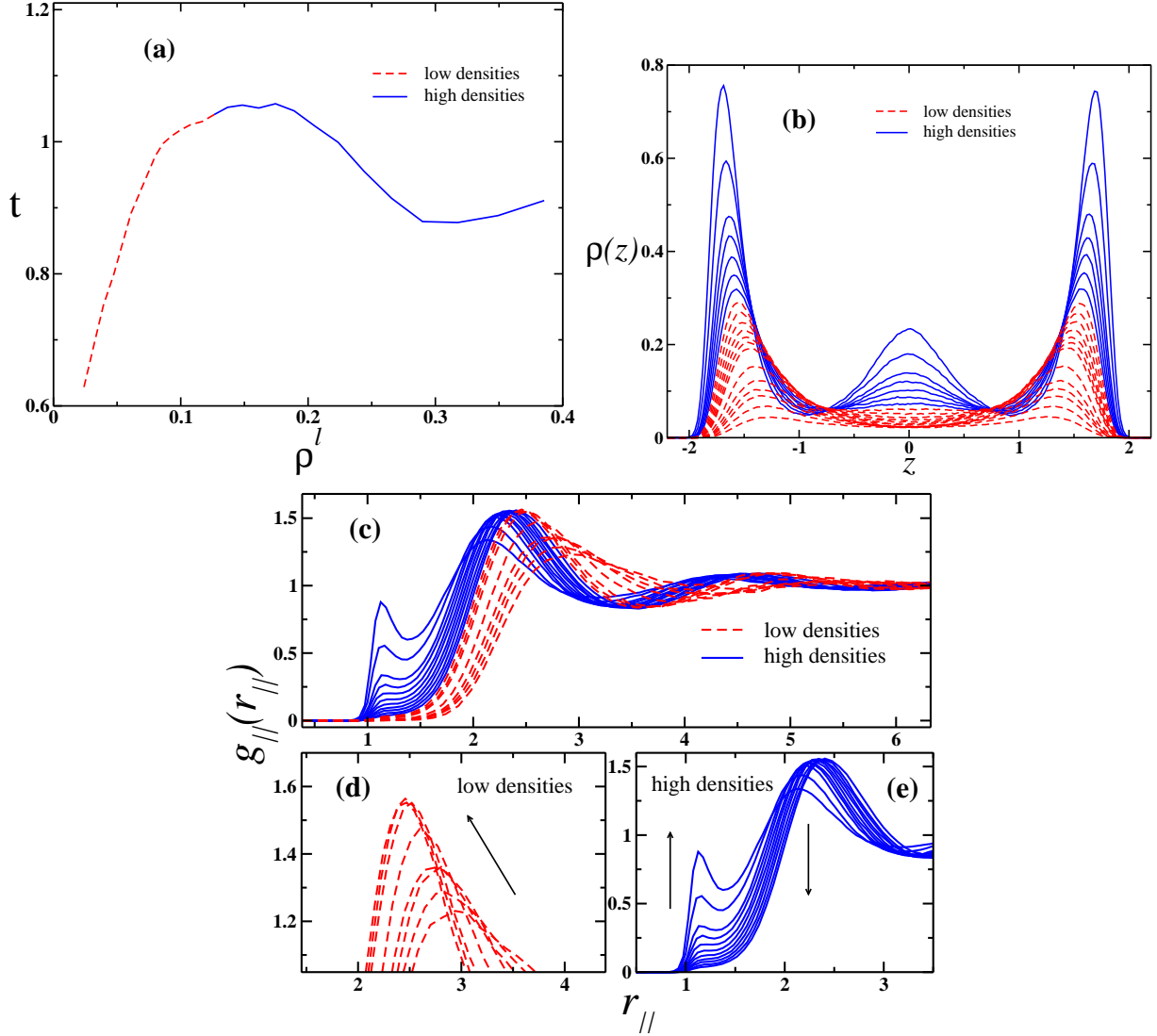

 \centering
 \begin{tabular}{cc}
\includegraphics[clip=true,width=8cm]{fig6a.eps}
\includegraphics[clip=true,width=8cm]{fig6b.eps}\\
\includegraphics[clip=true,width=10cm]{fig6c.eps}
\tabularnewline
 \end{tabular}\par
 \caption{Model A with plates separated by $L_z = 5.9$, temperature 
$T = 0.400$ and densities $\rho^l < 0.127$ (red dashed line) and 
$\rho^l \ge 0.127$ (blue solid line). In (a) we have the translational 
order parameter, in (b) the transversal density profile, in (c)
the lateral radial distribution function versus lateral distance, in (d) a
zoom of the firs peak of the LRDF for low densities and in (e) 
the competition between scales observed in the LRDF for high densities.
The arrows indicate the increase of density.}
\label{MA_t_400}
\end{figure}

The structure of the model A also was analyzed for high temperatures,
where the fluid shown only one region of structural anomaly. The 
Fig.~\ref{MA_t_400} (a) shows the translational oder parameter as 
function of layer density for plates separated by $L_z = 5.9$ and 
temperature $T = 0.400$. As we can see, for high temperatures, the higher 
entropic contribution for the fluid free energy leads the new structural 
anomaly to disappear. It occurs since the system changes from three layers at high densities
to a bulk-like profile at low densities, like shown in the transversal density profiles
in Fig.~\ref{MA_t_400} (b). Two well defined layers, like observed for very low densities at
$T = 0.150$ was not observed here, for $T = 0.400$. The lateral radial distribution 
function versus lateral distance was analyzed in Fig.~\ref{MA_t_400} (c). For 
low densities the system presents a bulk-like profile and the $g_{||}(r_{||})$ 
presents the first peak around $r_{||} \approx 2.5$. A zoom of this first 
peak is shown in Fig.~\ref{MA_t_400} (d). As the density increases, this first peak in the
$g_{||}(r_{||})$ increases, as indicated by the arrow, and consequently the parameter
$t$ increases. Whereas, for high densities, the first peak occurs at $r_{||} \approx 1.2$ 
and a competition between scales is observed (Fig.~\ref{MA_t_400} (e)) and the anomalous 
behavior in $t$ is detected. The two to three layers transition does not occur and, 
because that, the double region of structural anomaly is not present. The same 
behavior was observed for the model B in the cases of one or two anomalous region.
For simplicity, this results are not shown.

In the case of more attractive fluid-fluid potentials the
fluid-fluid interaction always wins against the wall-fluid
interaction and a middle layer that minimizes the fluid-fluid
interaction is always formed.

\begin{figure}[!htb]
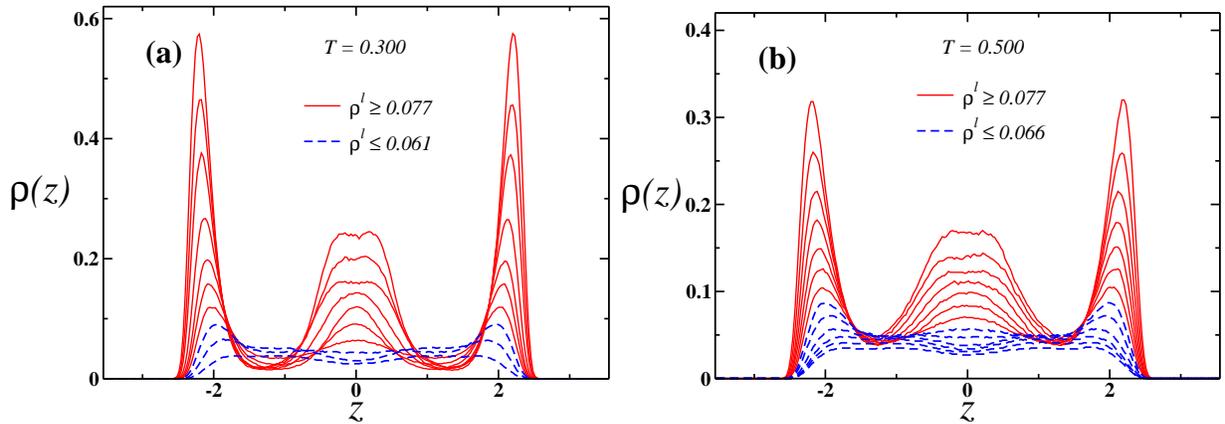

 \centering
 \begin{tabular}{ccc}
\includegraphics[clip=true,width=8cm]{fig7a.eps}
\includegraphics[clip=true,width=8cm]{fig7b.eps}
\tabularnewline
 \end{tabular}\par
 \caption{Transversal density profile for (a) model C at $T = 0.300$ and (b) model D at $T = 0.500$.
   Both systems were simulated with plates separated by $L_z = 7.1$.}
\label{histCD}
\end{figure}

Fig.~\ref{histCD} (a) and (b) show the transversal
density profiles for model C at $T = 0.300$ and model D
at $T = 0 .500$, respectively. In these models, the fluid changes
from three layers of particles to a bulk-like profile regardless the
system temperature. Therefore, the additional length that arises in
models A and B is not present, and the second region of anomaly in $t$ 
was not observed. The structural behavior for these models at low temperatures
is similar to what happens with the models A and B for high temperatures.

The anomalous behavior in translational order parameter is a well known
results for bulk systems~\cite{Tr00,Er01,Er03,Sh02,Er06,Ya05,Mi06a}. 
Similar to confined systems, Dudalov et al.~\cite{Du14a, Du14b, Du14c} 
analyzed the melting scenario of two-dimensional systems using structural 
order and found results very different from 3D cases. In the same way, 
our quasi-two-dimensional analysis also gives results completely different 
from the 3D bulk systems, arising the new anomaly caused by nanoconfinement.

\subsection*{Excess entropy}

\begin{figure}[!htb]
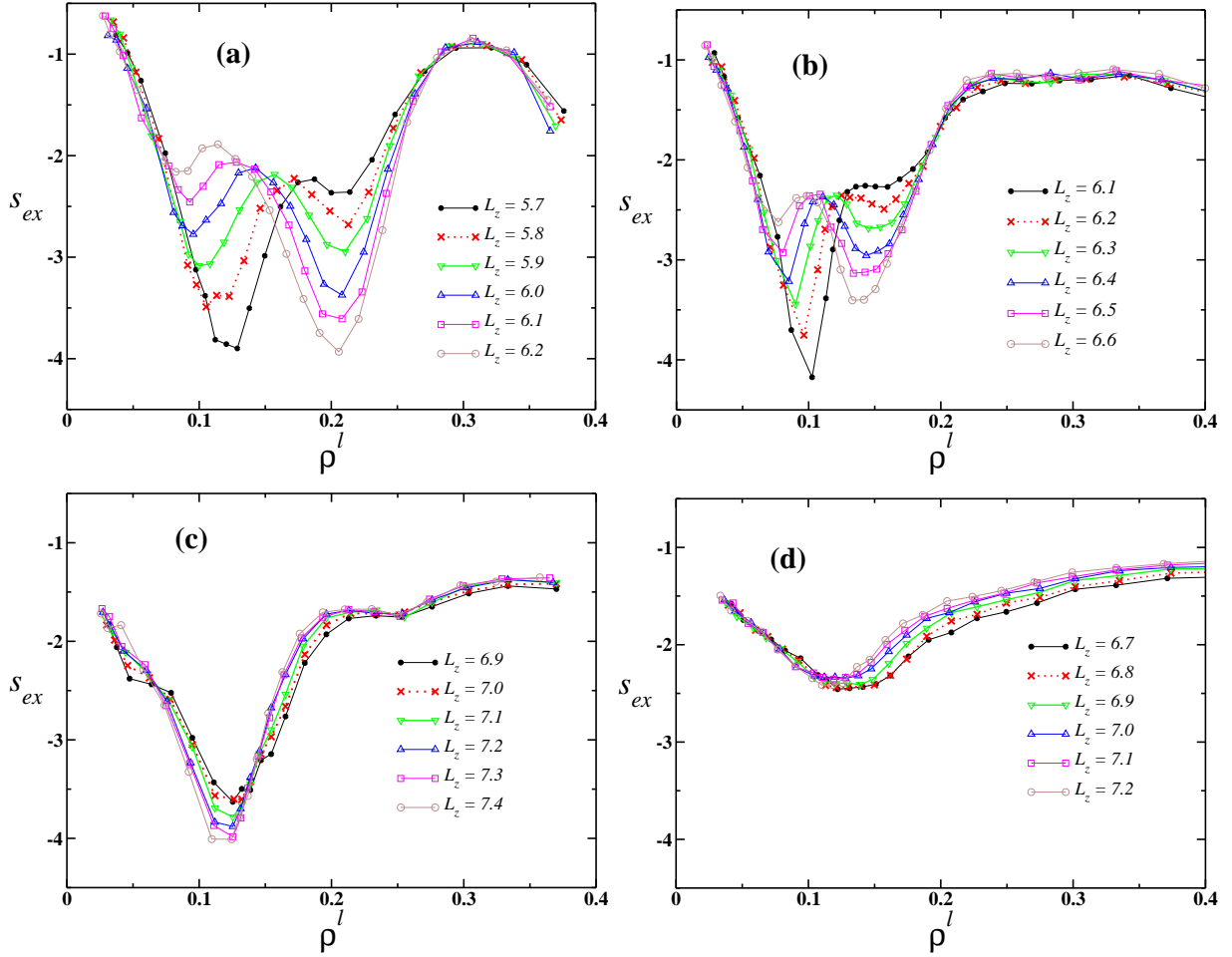

 \centering\begin{center}
 \begin{tabular}{cc}
\includegraphics[clip=true,width=8cm]{fig8a.eps}
\includegraphics[clip=true,width=8cm]{fig8b.eps}\\
\includegraphics[clip=true,width=8cm]{fig8c.eps}
\includegraphics[clip=true,width=8cm]{fig8d.eps}
\tabularnewline
 \end{tabular}\par
\end{center}
\caption{Excess entropy as function of layer density for (a)
model A at $T = 0.150$, (b) model B at $T = 0.200$, (c) model C
at $T = 0.300$ and (d) model D at $T = 0.500$. The other temperatures
and separation of plates were not shown for simplicity.}
\label{excess}
\end{figure}

A second measure that relates the structure with
the presence of anomalies is the excess entropy.
In order to confirm that the presence of 
the new anomalous region in 
the structure is also related to
anomalies in the thermodynamics, the 
excess entropy was computed for each of the models studied here. 
The models studied in this work in
the bulk exhibit the anomalous increase of $s_{ex}$ 
with the increase of density~\cite{Oliveira10,Silva10}. 

Fig.~\ref{excess} shows the excess entropy for the contact layer 
as function of layer density $\rho^l$ for (a) model A, (b) model B, 
(c) model C and (d) 
model D. The same behavior observed for the 
translational order parameter is seen for the excess 
entropy. For the models C and D the
same  anomalous region in the excess entropy
versus density phase diagram observed in bulk systems appears
under confinement. However, the models A and B exhibit an additional region
of anomaly for low temperatures and some distances $L_z$ between plates.

\begin{figure}[!htb]
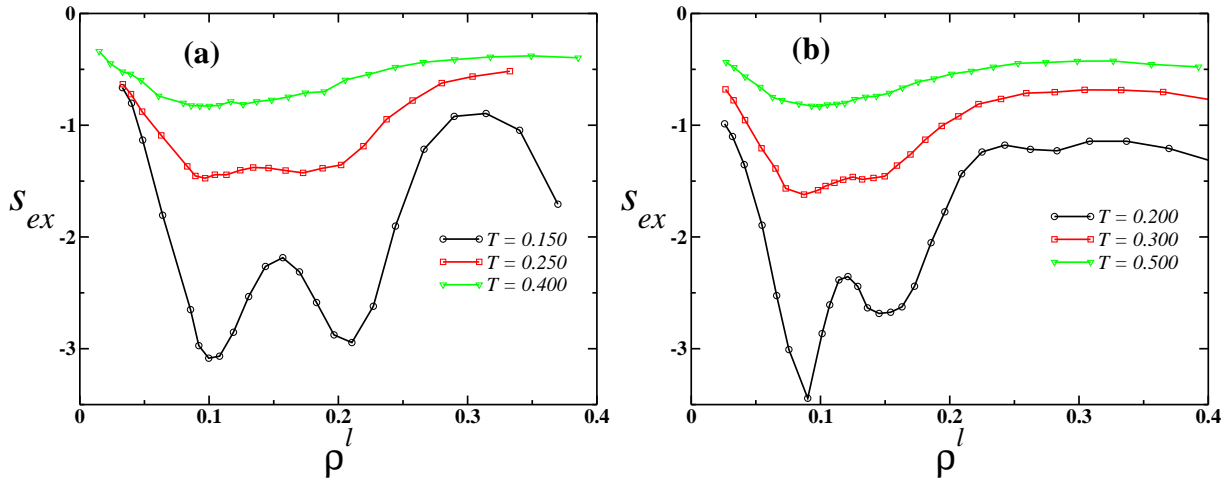

 \centering
 \begin{tabular}{ccc}
\includegraphics[clip=true,width=8cm]{fig9a.eps}
\includegraphics[clip=true,width=8cm]{fig9b.eps}
\tabularnewline
 \end{tabular}\par
 \caption{Excess entropy for (a) model A at $L_z = 5.9$ and (b) model B at $L_z = 6.3$.}
\label{EAB}
\end{figure}

The new region of anomaly in $s_{ex}$ is shown
in Fig.~\ref{EAB} (a) for the model A at $L_z = 5.9$ and $T = 0.150$
and (b) model B at $L_z = 6.3$ and $T = 0.200$. The densities
of maxima and minima of excess entropy, $\rho^l_{sexmax1},\rho^l_{sexmax},
\rho^l_{sexmin1}$ and $\rho^l_{sexmin}$, coincide with 
the densities of maxima and minima of translational
order parameter. This
shows that as the system changes 
from three to two layers the density of the 
contact layer increases, it becomes more structured 
and the entropy decreases what is 
a consistent picture. Similarly to
what happens in the $t$ behavior, as the temperature increases, the 
entropic effect leads the fluid to assume a bulk-like 
behavior and the new region of anomaly disappears.

The excess entropy is a good tool to understand the dynamic of bulk and 
confined fluids~\cite{Cho10,Bor12,Ing13,Sin10} and can be useful to see the presence
of density anomaly as well~\cite{Ba89,Er06,Cha06}. The increase of attractive 
well of core-softened models can destroy the water-like anomalies in 
bulk~\cite{Silva10} and the second region of anomaly in $t$ and $s_{ex}$ 
in confinement by plates.

\section{Conclusion}
\label{Conclu}

In this work we have analyzed the effects of confinement in a system
of particles interacting through models  of two 
length scale potential, varying from a  purely repulsive (model A) to
models that have
an attractive well (models B, C and D). 

We found that the confinement in the case of systems in which
the fluid-wall interaction competes with the fluid-fluid
interaction potentials leads to the appearance of 
new anomalous region in the translational order parameter
versus density phase diagram. This new anomalous behavior
is related to the change of structure that happens when
the system changes from three to two layers, namely when
the center layer melts.

The same behavior is observed from the thermodynamic side 
by the excess entropy that increases with increasing
density.

Our results indicates that by confining and particularly by
confining by system with strong interacting walls the 
confined fluid exhibits new phenomena not observed in 
the bulk systems.

\section{Acknowledgments}

We thanks the Brazilian agencies CNPq, INCT-FCx, and Capes for the financial support.


\begin{thebibliography}{10}

\bibitem{Ke75}
G.~S. Kell,
\newblock J. Chem. Eng. Data {\bf 20}, 97 (1975).

\bibitem{An76}
C.~A. Angell, E.~D. Finch, and P.~Bach,
\newblock J. Chem. Phys. {\bf 65}, 3065 (1976).

\bibitem{Pr87}
F.~X. Prielmeier, E.~W. Lang, R.~J. Speedy, and H.-D. L\"udemann,
\newblock Phys. Rev. Lett. {\bf 59}, 1128 (1987).

\bibitem{URL}
M.~Chaplin,
\newblock Sixty-nine anomalies of water,
\newblock \url{http://www.lsbu.ac.uk/water/anmlies.html}, 2013.

\bibitem{Th76}
H.~Thurn and J.~Ruska,
\newblock J. Non-Cryst. Solids {\bf 22}, 331 (1976).

\bibitem{Handbook}
{\em Handbook of Chemistry and Physics},
\newblock CRC Press, Boca Raton, Florida, 65 ed. edition edition, 1984.

\bibitem{Sa67}
G.~E. Sauer and L.~B. Borst,
\newblock Science {\bf 158}, 1567 (1967).

\bibitem{Ke83}
S.~J. Kennedy and J.~C. Wheeler,
\newblock J. Chem. Phys. {\bf 78}, 1523 (1983).

\bibitem{Ts91}
T.~Tsuchiya,
\newblock J. Phys. Soc. Jpn. {\bf 60}, 227 (1991).

\bibitem{Cu81}
P.~T. Cummings and G.~Stell,
\newblock Mol. Phys. {\bf 43}, 1267 (1981).

\bibitem{To97}
M.~Togaya,
\newblock Phys. Rev. Lett. {\bf 79}, 2474 (1997).

\bibitem{An00}
C.~A. Angell, R.~D. Bressel, M.~Hemmatti, E.~J. Sare, and J.~C. Tucker,
\newblock Phys. Chem. Chem. Phys. {\bf 2}, 1559 (2000).

\bibitem{Sh02}
M.~S. Shell, P.~G. Debenedetti, and A.~Z. Panagiotopoulos,
\newblock Phys. Rev. E {\bf 66}, 011202 (2002).

\bibitem{Sh06}
R.~Sharma, S.~N. Chakraborty, and C.~Chakravarty,
\newblock J. Chem. Phys. {\bf 125}, 204501 (2006).

\bibitem{Sa03}
S.~Sastry and C.~A. Angell,
\newblock Nature Mater. {\bf 2}, 739 (2003).

\bibitem{Ch06}
S.-H. Chen et~al.,
\newblock Proc. Natl. Acad. Sci. USA {\bf 103}, 12974 (2006).

\bibitem{Mo05}
T.~Morishita,
\newblock Phys. Rev. E {\bf 72}, 021201 (2005).

\bibitem{Ne01}
P.~A. Netz, F.~W. Starr, H.~E. Stanley, and M.~C. Barbosa,
\newblock J. Chem. Phys. {\bf 115}, 344 (2001).

\bibitem{Ne02a}
P.~A. Netz, F.~W. Starr, M.~C. Barbosa, and H.~E. Stanley,
\newblock Physica A {\bf 314}, 470 (2002).

\bibitem{Tr00}
T.~M. Truskett, S.~Torquato, and P.~G. Debenedetti,
\newblock Phys. Rev. E {\bf 62}, 993 (2000).

\bibitem{Er01}
J.~R. Errington and P.~D. Debenedetti,
\newblock Nature (London) {\bf 409}, 318 (2001).

\bibitem{Er03}
J.~E. Errington, P.~G. Debenedetti, and S.~Torquato,
\newblock J. Chem. Phys. {\bf 118}, 2256 (2003).

\bibitem{Ne58}
R.~E. Nettleton and H.~S. Green,
\newblock J. Chem. Phys. {\bf 29}, 1365 (1958).

\bibitem{Ra71}
H.~J. Ravech\'e,
\newblock J. Chem. Phys. {\bf 55}, 2242 (1971).

\bibitem{Wa87}
D.~C. Wallace,
\newblock J. Chem. Phys. {\bf 87}, 2282 (1987).

\bibitem{Ba89}
A.~Baranyai and D.~J. Evans,
\newblock Phys. Rev. A {\bf 40}, 3817 (1989).

\bibitem{Er06}
J.~R. Errington, T.~M. Truskett, and J.~Mittal,
\newblock J. Chem. Phys. {\bf 125}, 244502 (2006).

\bibitem{Tabeling14}
P.~Tabeling and L.~Bocquet,
\newblock Lab on a Chip {\bf 14}, 3143 (2014).

\bibitem{De10}
J.~Deschamps, F.~Audonnet, N.~Brodie-Linder, M.~Schoeffel, and
  C.~Alba-Simionesco,
\newblock Phys. Chem. Chem. Phys. {\bf 12}, 1440 (2010).

\bibitem{Ja08}
S.~J{\"a}hnert et~al.,
\newblock Phys. Chem. Chem. Phys. {\bf 10}, 6039 (2008).

\bibitem{Koga01}
K.~Koga, G.~T. Gao, H.~Tanaka, and X.~C. Zeng,
\newblock Nature {\bf 412}, 802 (2001).

\bibitem{MaG97}
M.~W. Maddox and K.~E. Gubbins,
\newblock J. Chem. Phys. {\bf 107}, 9659 (1997).

\bibitem{Giovambattista09}
N.~Giovambattista, P.~J. Rossky, and P.~G. Debenedetti,
\newblock Phys. Rev. Lett. {\bf 102}, 050603 (2009).

\bibitem{Lo09}
T.~G. Lombardo, P.~J. Rossky, and P.~G. Debenedetti,
\newblock Faraday Discuss. {\bf 141}, 359 (2009).

\bibitem{Zangi03a}
R.~Zangi and A.~E. Mark,
\newblock Phys. Rev. Lett. {\bf 91}, 025502 (2003).

\bibitem{Santos12}
F.~de~los Santos and G.~Franzese,
\newblock Phys. Rew. E {\bf 85}, 010602 (2012).

\bibitem{Du14a}
D.~E. Dudalov, Y.~D. Fomin, E.~N. Tsiok, and V.~N. Ryzhov,
\newblock Soft Matter {\bf 10}, 4966 (2014).

\bibitem{Du14b}
D.~E. Dudalov, Y.~D. Fomin, E.~N. Tsiok, and V.~N. Ryzhov,
\newblock Journal of Physics: Conference Series {\bf 510}, 012016 (2014).

\bibitem{Du14c}
D.~E. Dudalov, Y.~D. Fomin, E.~N. Tsiok, and V.~N. Ryzhov,
\newblock Physical Review Letters {\bf 112}, 157803 (2014).

\bibitem{Bordin14a}
J.~R. Bordin, L.~Krott, and M.~C. Barbosa,
\newblock J. Phys. Chem. C {\bf 118}, 9497 (2014).

\bibitem{Jakobtorweihen05}
S.~Jakobtorweihen, M.~G. Verbeek, C.~P. Lowe, .~F.~J.~Keil, and B.~Smit,
\newblock Phys. Rev. Lett. {\bf 95}, 044501 (2005).

\bibitem{Qin11}
X.~Qin, Q.~Yuan, Y.~Zhao, S.~Xie, and Z.~Liu,
\newblock Nanoletters {\bf 11}, 2173 (2011).

\bibitem{Bordin13a}
J.~R. Bordin, A.~Diehl, and M.~C. Barbosa,
\newblock J. Phys. Chem. B {\bf 117}, 7047 (2013).

\bibitem{Bordin14b}
J.~R. Bordin, J.~S. Soares, A.~Diehl, and M.~C. Barbosa,
\newblock J. Chem Phys. {\bf 140}, 194504 (2014).

\bibitem{Krott13b}
L.~Krott and J.~R. Bordin,
\newblock J. Chem. Phys. {\bf 139}, 154502 (2013).

\bibitem{Krott13a}
L.~Krott and M.~C. Barbosa,
\newblock J. Chem. Phys. {\bf 138}, 084505 (2013).

\bibitem{Bordin12b}
J.~R. Bordin, A.~B. de~Oliveira, A.~Diehl, and M.~C. Barbosa,
\newblock J. Chem. Phys {\bf 137}, 084504 (2012).

\bibitem{Ho81}
R.~G. Horn and J.~N. Israelachvili,
\newblock J. Chem. Phys {\bf 75}, 1400 (1981).

\bibitem{Gr91}
S.~Granick,
\newblock Science {\bf 253}, 1374 (1991).

\bibitem{Koga97}
K.~Koga, X.~C. Zeng, and H.~Tanaka,
\newblock Phys. Rev. Lett. {\bf 72}, 5262 (1997).

\bibitem{St12}
E.~G. Strekalova, J.~Luo, H.~E. Stanley, G.~Franzese, and S.~V. Buldyrev,
\newblock Phys. Rev. Lett. {\bf 109}, 105701 (2012).

\bibitem{Fa09}
A.~Faraone, K.-H. Liu, C.-Y. Mou, Y.~Zhang, and S.-H. Chen,
\newblock J. Chem. Phys. {\bf 130}, 134512 (2009).

\bibitem{Er11}
M.~Erko, N.~Cade, A.~G. Michette, G.~H. Findenegg, and O.~Paris,
\newblock Phys. Rev. B {\bf 84}, 104205 (2011).

\bibitem{Krott14a}
L.~Krott and M.~C. Barbosa,
\newblock Phys. Rev. E {\bf 89}, 012110 (2014).

\bibitem{Oliveira06a}
A.~B. de~Oliveira, P.~A. Netz, T.~Colla, and M.~C. Barbosa,
\newblock J. Chem. Phys. {\bf 124}, 084505 (2006).

\bibitem{Oliveira06b}
A.~B. de~Oliveira, P.~A. Netz, T.~Colla, and M.~C. Barbosa,
\newblock J. Chem. Phys. {\bf 125}, 124503 (2006).

\bibitem{Silva10}
J.~N. da~Silva, E.~Salcedo, A.~B. de~Oliveira, and M.~C. Barbosa,
\newblock J. Chem. Phys. {\bf 133}, 244506 (2010).

\bibitem{AllenTild}
P.~Allen and D.~J. Tildesley,
\newblock {\em Computer Simulation of Liquids},
\newblock Oxford University Press, Oxford, 1987.

\bibitem{Ku05}
P.~Kumar, S.~V. Buldyrev, F.~Sciortino, E.~Zaccarelli, and H.~E. Stanley,
\newblock Phys. Rev. E {\bf 72}, 021501 (2005).

\bibitem{kumar07}
P.~Kumar, F.~W. Starr, S.~V. Buldyrev, and H.~E. Stanley,
\newblock Phys. Rev. E {\bf 72}, 011202 (2007).

\bibitem{Ya05}
Z.~Yan, S.~V. Buldyrev, N.~Giovambattista, and H.~E. Stanley,
\newblock Phys. Rev. Lett. {\bf 95}, 130604 (2005).

\bibitem{Mi06a}
J.~Mittal, J.~R. Errington, and T.~M. Truskett,
\newblock J. Phys. Chem. B {\bf 110}, 18147 (2006).

\bibitem{Oliveira10}
A.~B. de~Oliveira, E.~Salcedo, C.~Chakravarty, and M.~C. Barbosa,
\newblock J. Chem. Phys. {\bf 132}, 234509 (2010).

\bibitem{Cho10}
R.~Chopra, T.~M. Truskett, and J.~R. Errington,
\newblock Phys. Rev. E {\bf 82}, 041201 (2010).

\bibitem{Bor12}
B.~J. Borah, P.~K. Maiti, C.~Chakravarty, and S.~Yashonath,
\newblock J. Chem. Phys. {\bf 136}, 174510 (2012).

\bibitem{Ing13}
T.~S. Ingebrigtsen, J.~R. E. T.~M. Truskett, and J.~C. Dyre,
\newblock Phys. Rev. Lett. {\bf 111}, 235901 (2013).

\bibitem{Sin10}
M.~Singh, H.~Liu, S.~K. Kumar, A.~Ganguly, and C.~Chakravarty,
\newblock J. Chem. Phys. {\bf 132}, 074503 (2010).

\bibitem{Cha06}
S.~N. Chakraborty and C.~Chakravarty,
\newblock J. Chem. Phys. {\bf 124}, 014507 (2006).

\end{thebibliography}

\end{document}